\documentclass[a4paper,11pt]{article}
\pdfoutput=1 
             
\usepackage{lineno}
\usepackage{subfigure}
\usepackage{jinstpub} 

\title{\boldmath Radiation hard DMAPS pixel sensors in 150\,nm CMOS technology for operation at LHC}

\author[b]{M. Barbero,}
\author[b]{P. Barrillon,} 
\author[a]{C. Bespin,} 
\author[b]{S. Bhat,}
\author[b]{P. Breugnon,} 
\author[a]{I. Caicedo,} 
\author[b]{Z. Chen,}
\author[c]{Y. Degerli,}
\author[a]{J. Dingfelder,}
\author[b]{S. Godiot,} 
\author[c]{F. Guilloux,}
\author[a]{T. Hemperek,} 
\author[a,1]{T. Hirono,\note{Corresponding authors.}}
\author[a]{F. H\"ugging,} 
\author[a]{H. Kr\"uger,} 
\author[a]{K. Moustakas,}
\author[c]{A. Ouraou,} 
\author[b]{P. Pangaud,} 
\author[d]{I. Peric,} 
\author[a]{D-L. Pohl,} 
\author[a]{P. Rymaszewski,} 
\author[c]{P. Schwemling,} 
\author[c]{M. Vandenbroucke,}
\author[a,1]{T. Wang,}
\author[a]{N. Wermes}

\affiliation[a]{University of Bonn, Physikalisches Institut, Nu{\ss}allee 12, 53115 Bonn, Germany}
\affiliation[b]{Aix Marseille University, CNRS/IN2P3, Centre de Physique des Particules de Marseille, 163 Avenue de Luminy, Marseille, France}
\affiliation[c]{IRFU, CEA-Saclay, Gif-sur-Yvette Cedex, 91191 France}
\affiliation[d]{Karlsruhe Institute for Technology, 76131 Karlsruhe, Germany}

\emailAdd{hirono@physik.uni-bonn.de}
\emailAdd{t.wang@physik.uni-bonn.de}

\abstract{Monolithic Active Pixel Sensors (MAPS) have been developed since the late 1990s employing silicon substrate with a thin epitaxial layer in which deposited charge is collected by disordered diffusion rather than by drift in an electric field. As a consequence the signal is small and slow, and the radiation tolerance is below the requirements for LHC experiments by factors of 100 to 1000.
We developed fully depleted (D)MAPS pixel sensors employing a 150\,nm CMOS technology and using a high resistivity substrate as well as a high biasing voltage. The development has been carried out in three subsequent iterations, from prototypes to a large pixel matrix comprising a complete readout architecture suitable for LHC operation. Full CMOS electronics is
embedded in large deep n-wells which at the same time serve as collection nodes (large electrode design). 
The devices have been intensively characterized before and after irradiation employing lab tests as well as particle beams. The devices can cope with particle rates seen by the innermost pixel detectors
of the LHC pp-experiments or as seen by the outer pixel layers of the planned HL-LHC upgrade. They are radiation hard to particle fluences of at least $10^{15}~\mathrm{n_{eq}/cm^2}$ and total ionization doses of at least 50\,Mrad.}


\arxivnumber{1911.01119} 

\begin{document}
\maketitle
\flushbottom
\section{Introduction}
\label{sec:intr}
The current state of the art of pixel detectors used in high energy particle physics experiments are so-called \emph{hybrid pixels} where the depleted sensing part, a pixel diode structure, and the readout chip are different entities, mated using the bump-bonding and flip-chipping technology connecting each pixel of both parts \cite{pixel_book}. 
Tuning of the sensing and the signal processing tasks can be done for both parts separately, hence constituting an advantage and rendering their use possible in high rate and radiation experiments, as encountered for example at the LHC (see for example \cite{Garcia-Sciveres:2017ymt} for a recent review). 
Hybrid pixels also have some notable disadvantages, however, the main one 
being the labor- and cost-intensive assembly of modules. Also the 
material budget of modules is comparatively large.

The principle of monolithic pixel detectors for particle detection, where the sensing and the readout electronics parts form one (monolithic) entity, was first realized as early as 1994 \cite{Snoeys1994}. The use of CMOS technologies for monolithic pixel modules has started in the late 1990ies \cite{Dierickx1998,Turchetta_2001}, first exploiting a thin (typ.\,20\,$\upmu$m) epitaxial layer, often offered in CMOS imaging technologies. These devices, however, do not withstand the high radiation levels
at the LHC. Bulk depletion of CMOS based pixel detectors was achieved employing a high voltage (HV) process \cite{Peric_2007}. We have addressed the goal to achieve fully depleted monolithic pixel sensors (DMAPS) using high ohmic bulk material (>\,2 k$\Omega$\,cm) and high biasing voltages (>\,200\,V) for many years,
prototyped in different technologies  
\cite{Havranek:2014ora,Hemperek:2014yoa,Hirono_2016_IEEE,Wang_2017,Wang_2018,Hirono_2018,Moustakas_2018,Caicedo:2019lrk}.
The 150\,nm CMOS process offered by LFoundry \cite{LFoundry} was found to be well suited to satisfy the demands imposed by the particle environment encountered at the LHC pp experiments in terms of rate and radiation.  

In this paper, we present the results of the design, the development, and the characterization of fully depleted, radiation hard, and high rate capable  DMAPS sensors in a 150\,nm CMOS technology. The development has been prototyped in a sequence of three generations since 2014, from small pixel matrices with simplified readout to large (1$\times$1\,cm$^2$) matrices with full readout architecture, suitable to operate in rate/radiation environments as existing at the innermost pixel layers of current LHC pp experiments or at outer layers in a HL-LHC upgrade scenario, that is fluences around $10^{15}\mathrm{n_{eq}/cm^2}$, total ionising doses around 50--100 Mrad and particle rates of 100--200 MHz/cm$^2$ \cite{hl-lhc-tdr}.  
Full CMOS functionallity has been maintained, i.e. equal and unconstrained usage of PMOS and NMOS transistors. 

Section \ref{sec:lfdev} gives an overview of the LFoundry 150\,nm technology and the prototype devices used. Section \ref{sec:design} presents the design of the LF-Monopix1 chip with complete readout architecture. In section \ref{sec:res} the carried out characterisations are described, including test beam results of irradiated devices. The paper is concluded in section \ref{sec:con}.

\section{Overview of DMAPS development in the LFoundry 150~nm CMOS process}
\label{sec:lfdev}


\subsection{Sensor design concept}
\label{sec:lfdev_concept}
\begin{figure}
  \centering
  \includegraphics[width=.95\textwidth]{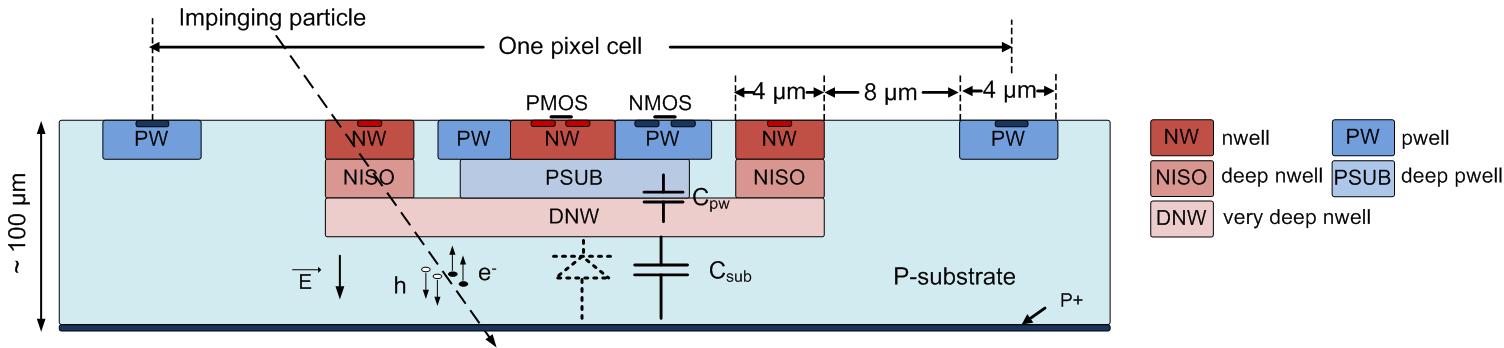}
  \caption{Schematic cross-section of a pixel cell implemented in the LFoundry 150~nm CMOS technology. Red color represents n-type material and blue color represents p-type material. The charges created in the p-substrate are collected by a n-type collection electrode formed by a very deep n-well implant (DNW). A deep p-well (PSUB) is used to shield the n-well (NW) hosting the PMOS transistor from the charge collection node. The p-wells (PW) on both sides of DNW are p-stop implants to isolate two neighboring pixels. The drawing is not to scale and only illustrates the relative implantation depths of different wells. In reality, the dimensions of the well structures are negligible in depth compared to the sensitive volume (p-substrate).}
  \label{fig:crosssection}
\end{figure}

Figure~\ref{fig:crosssection} depicts the cross section of a pixel cell. Unlike the conventional MAPS structure as for example the one described in~\cite{Turchetta_2001}, which employs a small n-well in a p-type epitaxial layer as the charge collection diode, the proposed design uses a large deep buried n-well (DNW) in a p-type wafer substrate for this task. As a matter of fact, such sensor implementation with a large collection electrode is intrinsically similar to the standard silicon n-in-p planar pixel sensor employed for LHC hybrid pixels in ATLAS or CMS~\cite{Gallrapp_2012}. 
A high reverse bias voltage is applied to the charge collection diode, such that a thick depleted region with a strong directing electric field is created in the substrate. Therefore, the charges created in the depleted region by impinging ionizing particles drift along the field lines to the charge collection node. This ensures fast charge collection and reduces the probability of signal charge being trapped in defect energy levels created by irradiation. 
 
The electronic devices can be implemented within the volume above the DNW, located in their respective wells, i.e. NMOS transistors in n-wells (NW) and PMOS transistors in p-wells (PW). The electronic layer is fully enclosed by the n-well structures (DNW, NISO and NW). It is isolated from the p-substrate and operates in its own low voltage domain (i.e.\,1.8\,V). A similar design was first proposed in~\cite{Peric_2007}, but the work presented in this paper enhances the sensor capability in two aspects:

\begin{enumerate}
  \item A deep p-well (PSUB) is available in the chosen technology, which is used as a shielding layer between the n-wells of the PMOS transistors and the charge collecting DNW. Therefore, PMOS transistors can be used within the pixel without introducing parasitic charge collection paths, resulting in more flexibility and more options for the in-pixel circuit design (full CMOS).
  \item A high-resistivity wafer qualified by the foundry is used. The quoted resistivity of the wafer substrate is larger than 2~k$\Omega \cdot$cm, 
  while the resistivity of standard wafer material typically is only in the range of ~$\sim$\,10~$\Omega \cdot$cm. Using high resistivity substrate and large biasing voltage allows for a large depleted region, for which the approximate dependence is
  $d \propto \sqrt{\rho\, V}$ (for a junction topology), where $\rho$ is the resistivity of substrate and $V$ the bias voltage.
\end{enumerate}

With proper sensor guard ring design, a reverse bias voltage higher than 200\,V can generally be applied. Moreover, wafer thinning and backside implantation are possible after the CMOS processing. Thus, a fully depleted and thin sensor, for example in the order of 100\,$\upmu$m, biased from the backside can be achieved, providing a strong electric field for uniform charge collection by drift inside the sensor. 

\subsection{Design challenges}
\label{sec:lfdev_challenges}

The implementation of a fully monolithic sensor with complex integrated in-pixel circuitry imposes some design challenges. As shown in figure~\ref{fig:crosssection}, the junction capacitance between the DNW and its inner PSUB/PW implants (C$_{\text{pw}}$) contributes to the total sensor capacitance. The value of the capacitance per area of C$_{\text{pw}}$ is generally high due to the high doping concentrations of the wells and the small bias voltage. Simulations show that the capacitance per unit area is $\sim$~0.11~fF/$\upmu$m$^{2}$ for the DNW-PSUB junction at a typical bias voltage of 2\,V, and slightly smaller for the DNW-PW junction due to the larger distance between them~\cite{Hemperek_Thesis}. 
Note that the PSUB/PW volume is the ``bulk'' for the in-pixel electronics, and the contribution of C$_{\text{pw}}$ to the total detector the capacitance can be significant when a large PSUB/PW area is needed to implement complex in-pixel digital electronics. The increase of detector capacitance requires more power for the analog front-end circuit in order to maintain the same timing performance, and leads to larger noise~\cite{pixel_book}. Moreover, placing digital electronics inside a sensitive collection node calls for very careful circuit design in order to minimize any potential cross talk from the digital transients to the sensing node. In particular, ``substrate noise'' introduced by the digital circuit is directly coupled to the sensing node via C$_{\text{pw}}$. Therefore the potential of the PSUB/PW region needs to be as stable as possible. Otherwise, the minimal operational threshold may be adversely impacted. The measures taken
are described further in sections \ref{sec:PixelLogic} and \ref{sec:Layout}.


\subsection{Development line towards a fully monolithic pixel sensor matrix}
\label{sec:lfdev_devline}

The design of our first fully monolithic prototype, LF-Monopix1, follows the development of its two predecessors: CCPD\_LF and LF-CPIX. The main parameters for the three prototypes in the family are listed in table~\ref{tab:LF_Chips}. 

\paragraph{CCPD\_LF}
CCPD\_LF is a small-scale, active pixel-sensor chip that is an embodiment of the ``smart diode'' concept~\cite{Peric_2012}. With a preamplifier and discriminator circuit equipped in each pixel, the monolithic configuration of CCPD\_LF can have a frame-based readout. The fast data-driven readout can be achieved by hybridization with a dedicated readout chip, either via the conventional bump-bonding process or using the capacitive coupling concept~\cite{Peric_2009}. Full characterization has been performed for CCPD\_LF, showing good radiation tolerance up to 50~Mrad total ionization dose (TID) and to particle fluences of $1 \times 10^{15}~\text{n}_{\text{eq}}$/$\text{cm}^{2}$ ~\cite{Hirono_2016_IEEE,Hirono_2016_NIM}.

\paragraph{LF-CPIX}
LF-CPIX is a large scale demonstrator chip that includes electronics and sensor guard ring improvements with respect to CCPD\_LF~\cite{Degerli_2016,Liu_2017}. The pixel size is 250~$\upmu$m\,$\times $\,50$\upmu$m, which is the same pixel size as for the ATLAS FE-I4 chip~\cite{Garcia_2011}. As compared to CCPD\_LF, improved sensor breakdown behavior and TID tolerance have been measured for the LF-CPIX chip~\cite{Hirono_2018}. The detection efficiency of an un-irradiated chip was measured in a beam test to be $\sim$\,99.5\,\%, and was not influenced by sensor thinning and back side processing~\cite{Hirono_Thesis}.
\paragraph{LF-Monopix1}
Inheriting from the LF-CPIX design, the LF-Monopix1 chip has greatly enhanced the capability of pixel-level digital processing~\cite{Wang_2017,Rymazewski_2018}. 
It incorporates the so called \textbf{\textit{column drain}} readout architecture, which was established for hybrid pixel readout chips and has been used by the current ATLAS pixel detector~\cite{Einsweiler_1999,Mandelli_2002,Peric_2006_FEI3}. The expected rate capability for a full scale sensor chip (i.e. 2~cm\,$\times $\,2~cm) is higher than 100~MHz/cm$^{2}$ (depending on 
hit topologies), hence meeting the requirements of either the current inner layers (e.g. the ATLAS B-layer) of LHC pp-experiments or the outer pixel layers for HL-LHC upgrade experiments~\cite{Arutinov_2009}. The goal of LF-Monopix1 was to address the design challenges presented in~section~\ref{sec:lfdev_challenges} and to demonstrate the feasibility of a large monolithic column drain readout matrix with complex in-pixel digital circuitry.

\begin{table}[htbp]
    \centering
    \caption{Parameters of prototype chips in the LFoundry DMAPS development line}
    \label{tab:LF_Chips}
    \smallskip
    \begin{tabular}[t]{l|c|c|c}
        \hline
        & CCPD\_LF & LF-CPIX & LF-Monopix1 \\
        \hline
        Chip size  & 5~mm $\times$ 5~mm & 10~mm $\times$ 10~mm & 10~mm $\times$ 10~mm \\ 
        \hline
        Pixel size & 33.3~$\mathrm{\upmu}$m $\times$ 125~$\mathrm{\upmu}$m & 50~$\mathrm{\upmu}$m $\times$ 250~$\mathrm{\upmu}$m  &  50~$\mathrm{\upmu}$m $\times$ 250~$\mathrm{\upmu}$m \\
        \hline
        Matrix size & 114 rows $\times$ 24 columns & 158 rows $\times$ 36 columns & 129 rows $\times$ 36 columns\\
        \hline
        Readout mode  & \multicolumn{2}{c|}{Frame based R/O}  & Fast column drain R/O \\
          & \multicolumn{2}{c|}{or fast R/O with dedicated R/O chip}  &  \\
        \hline
        Tape out & Sep. 2014 & Feb. 2016 & Aug. 2016     \\
        \hline
    \end{tabular}
\end{table}

\section{Design of LF-Monopix1}
\label{sec:design}


\subsection{Pixel design}
\label{sec:pixel}

The pixel size of LF-Monopix1 is the same as the LF-CPIX pixel, i.e. 50\,$\times$\,\,250\,$\upmu$m$^{\text{2}}$. Its analog front-end circuit is similar to LF-CPIX, but optimized to deal with the larger sensor capacitance $\sim$\,400\,fF expected for LF-Monopix1. The big difference in design of LF-Monopix1, as compared to its ancestors, is the inclusion of digital readout logic in the pixel. In the following sections, the details of this in-pixel design are presented.

\subsubsection{Analog front-end circuit}
\label{sec:FE}

\begin{figure}[htbp]
  \centering
  \includegraphics[width=.7\textwidth]{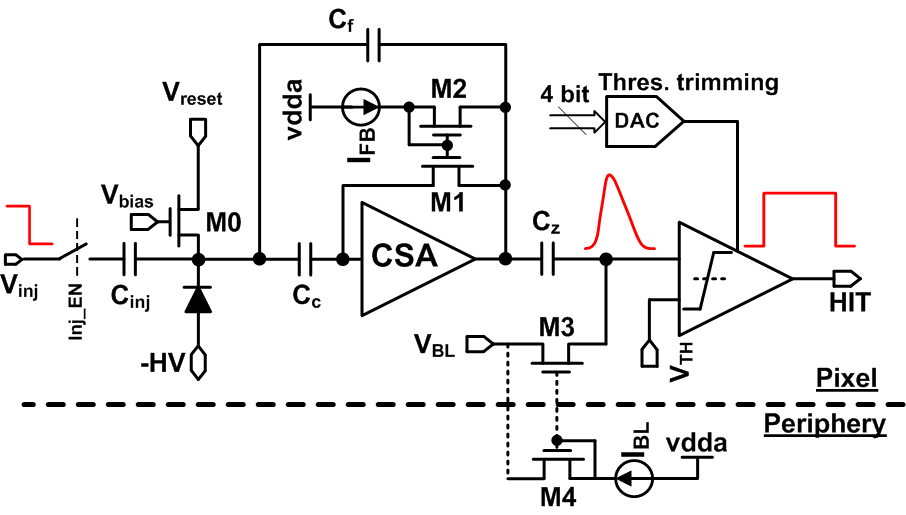}
  \caption{Circuit diagram of the analog front-end circuit.}
  \label{fig:FE}
\end{figure}

The schematic of the analog front-end (AFE) circuit is shown in figure\,\ref{fig:FE}. The sensing diode is biased with an NMOS transistor. The capacitor C$_{\text{c}}$, implemented with PMOS transistors, decouples the sensor leakage path from the preamplifying stage and therefore no leakage compensation circuit is needed. The preamplifying stage is a charge sensitive amplifier (CSA) with the input charge signal integrated on the feedback capacitor C$_{\text{f}}$. The DC feedback, mainly composed of a current mirror (M1 and M2), stabilizes the operation point and continuously discharges the feedback capacitor. The bias current I$_{\text{FB}}$ controls the discharge of the feedback capacitor, and is adjustable by a global DAC (Digital-to-Analog Converter). Such a DC feedback allows for high feedback resistance, regardless of the discharge current \cite{Blanquart_1997}. The voltage pulse generated by the CSA is sent to a discriminator circuit, which fires when the analog pulse crosses a pre-defined threshold V$_{\text{TH}}$. The discriminator output holds until the analog pulse falls again below this threshold. A 4-bit current DAC is implemented in each pixel to trim the discriminator threshold and thus minimize the threshold dispersion between readout channels. The capacitor C$_{\text{z}}$ decouples the operational point of CSA and discriminator. The baseline of the discriminator input is set to V$_{\text{BL}}$ via the MOS resistor M3. The biasing of the MOS resistor M3 is set globally by a DAC adjusting the current I$_{\text{BL}}$. For calibration and characterization purposes, a charge injection circuit is included. By applying a negative voltage (< 1.2\,V) pulse to the capacitor C$_{\text{inj}}$ ($\approx$\,2\,fF) a corresponding charge is injected into the amplifier.
The total static current consumption for the analog front-end circuit is $\sim$\,20\,$\upmu$A. In this prototype, both the preamplifier and the discriminator have two design variants which are described in the following paragraphs.

\begin{figure}[!ht]
  \begin{center}
    \subfigure[]{
      \includegraphics[width=0.21\textwidth]{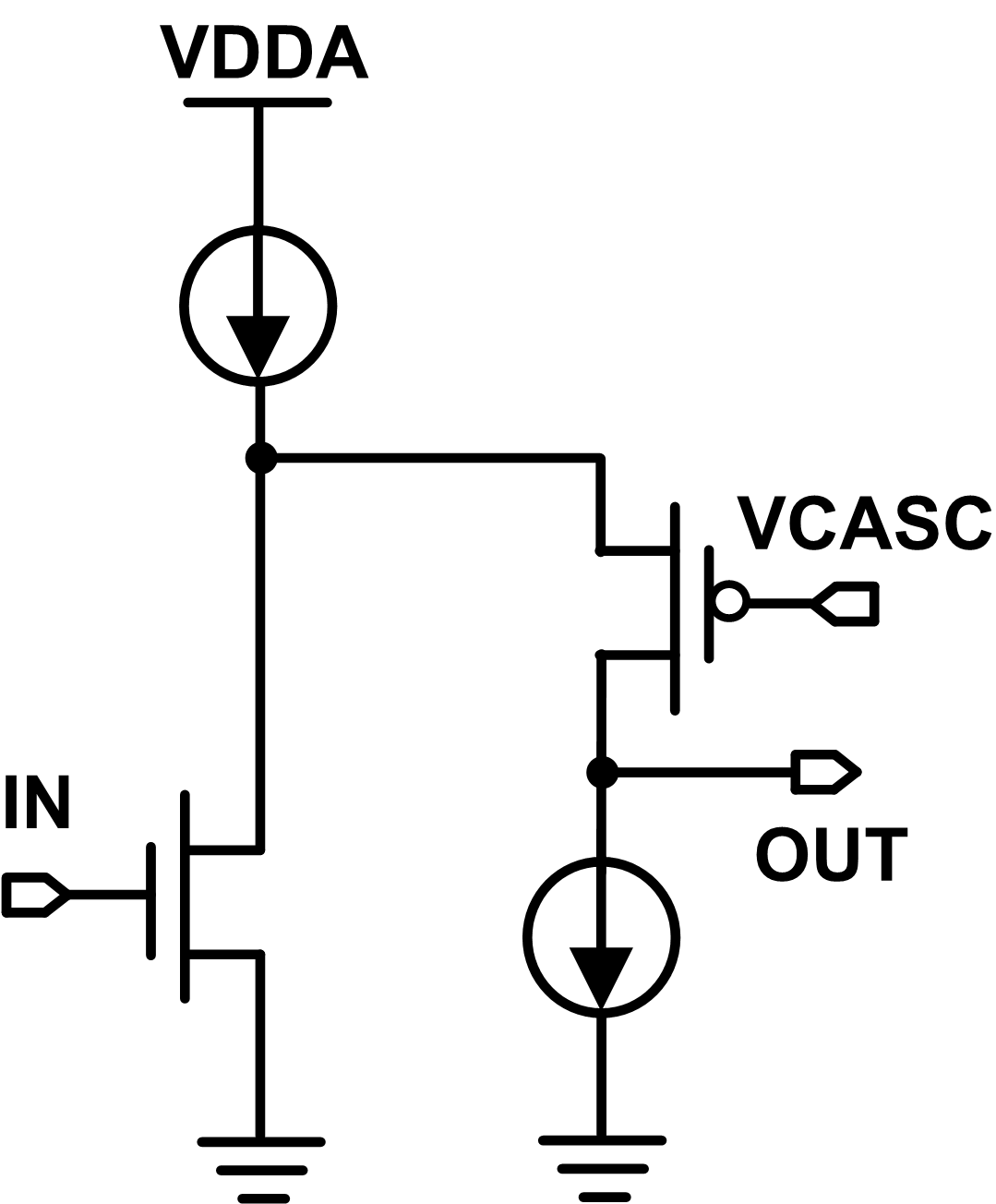}
      \label{fig:NMOS}
    }
    \qquad
    \qquad
    \subfigure[]{
      \includegraphics[width=0.22\textwidth]{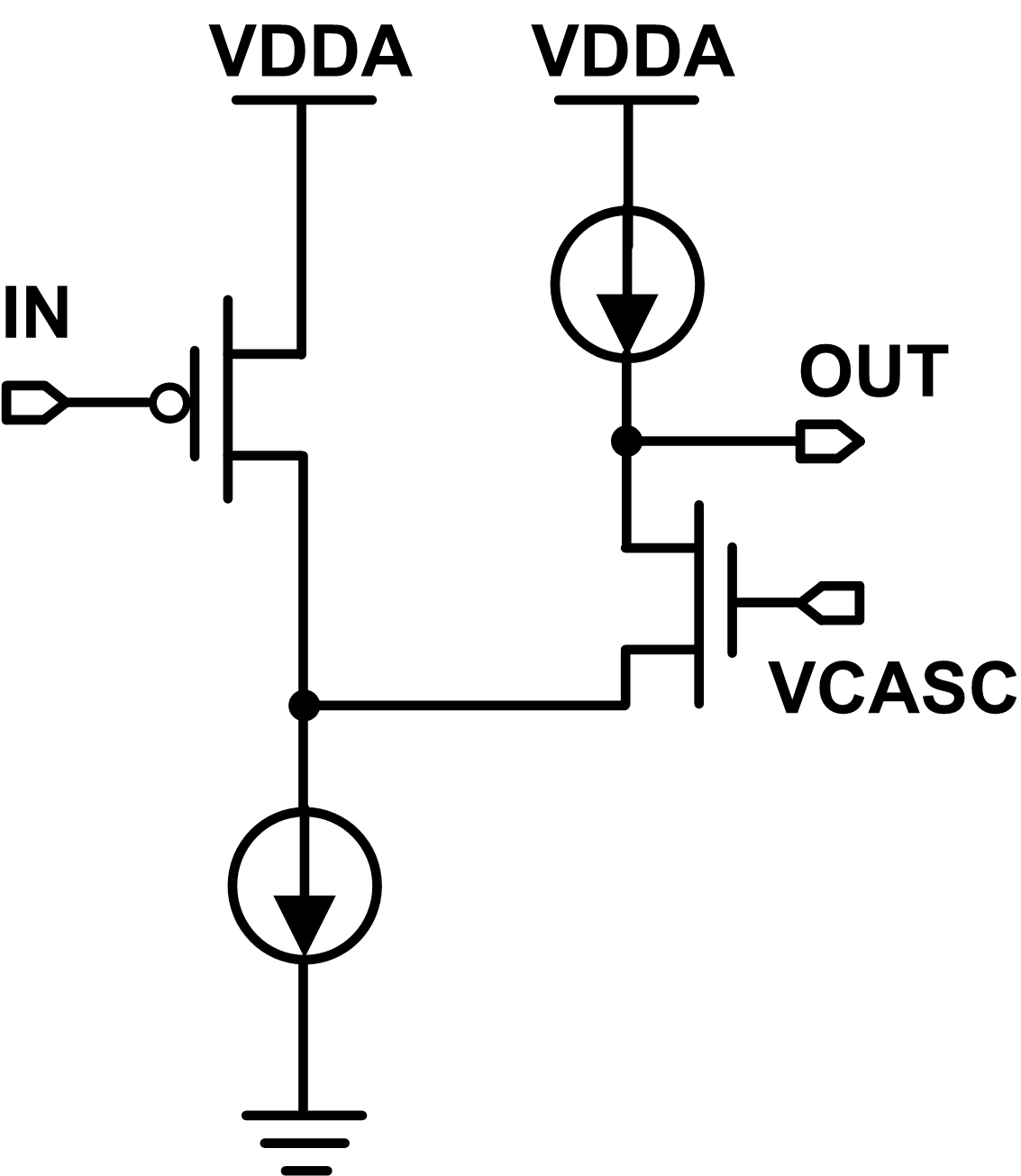}
      \label{fig:PMOS}
    }
    \qquad
    \qquad
    \subfigure[]{
      \includegraphics[width=0.25\textwidth]{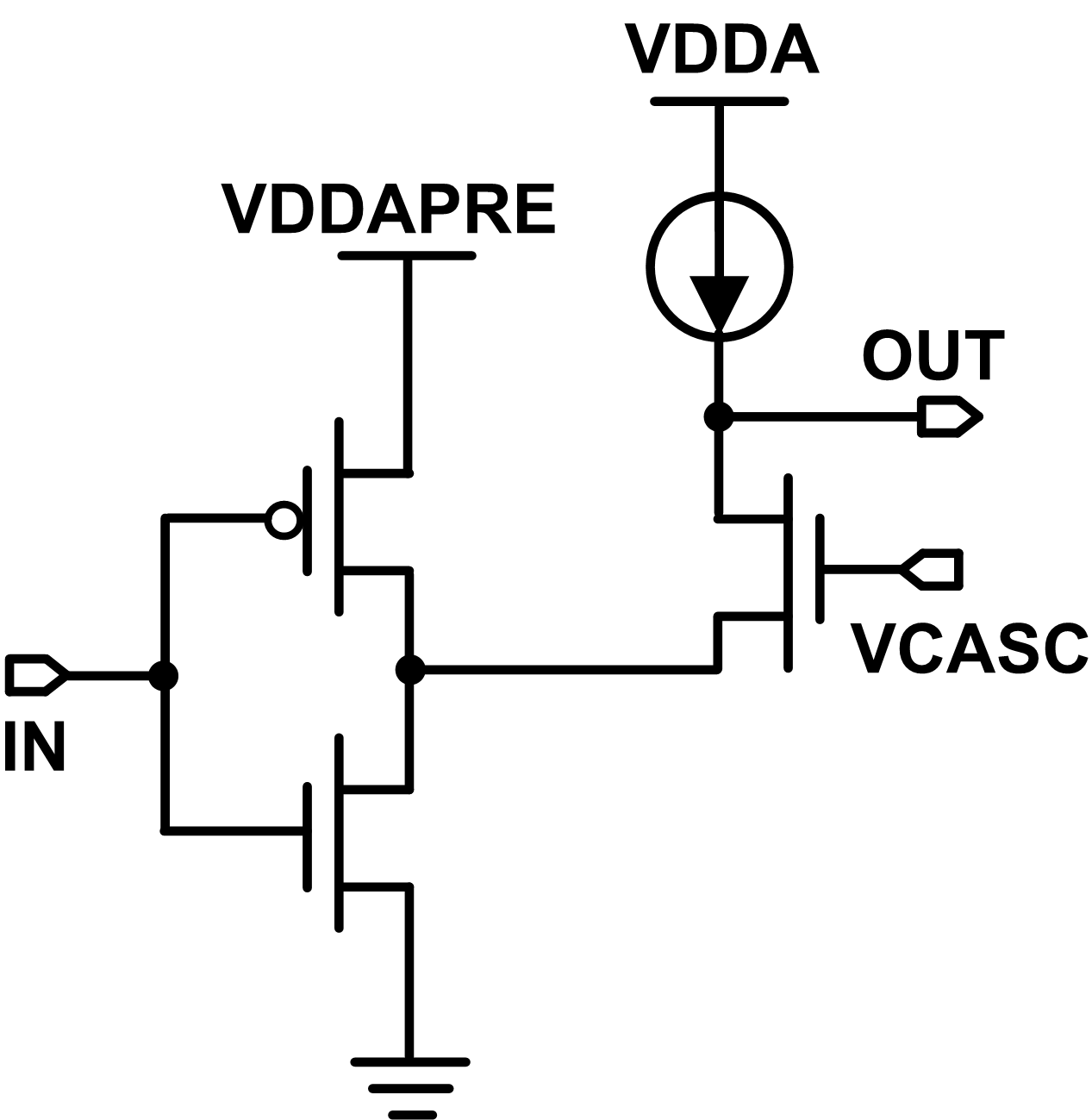}
      \label{fig:CMOS}
    }
    \caption[]{Schematics of three preamplifier designs using \subref{fig:NMOS} an NMOS transistor as the input device, \subref{fig:PMOS} a PMOS input transistor, and \subref{fig:CMOS} both NMOS and PMOS as input transistors, repectively. All three designs are implemented in LF-CPIX, while only \subref{fig:NMOS} and \subref{fig:CMOS} are implemented in LF-Monopix1.} 
    \label{fig:PreAmp}
  \end{center}
\end{figure}

\paragraph{Preamplifier} The schematics of three preamplifier variants tested are shown in figure~\ref{fig:PreAmp}. The amplifiers in figures \ref{fig:NMOS} and \ref{fig:PMOS} are typical single-stage folded cascode amplifiers using different input transistor types, while the one in figure\,\ref{fig:CMOS} makes use of both transistor types as input devices. The latter has an increased transconductance for a given bias current compared to the former ones, and its bias current is controlled by a separate analog power VDDAPRE regulated at the periphery\,\cite{Degerli_2016}. The CSA using the amplifier in figure~\ref{fig:CMOS} is later referred to as CMOS-CSA. All the three amplifier variants have been implemented in the former LF-CPIX chip, and the nominal bias current is 14\,$\upmu$A. However, only variants in figure~\ref{fig:NMOS} and figure~\ref{fig:CMOS} have been realized in the LF-Monopix1 chip. In order to cope with the larger sensor capacitance ($\sim$\,400\,fF) in the LF-Monopix1 chip due to the additional in-pixel digital circuitry with respect to the LF-CPIX chip, the nominal bias current has been increased to 17.5\,$\upmu$A for the NMOS input amplifier and 15\,$\upmu$A for the amplifier in figure\,\ref{fig:CMOS}, bearing in mind a fast desired peaking time of $\sim$\,25\,ns.

\begin{figure}[!ht]
  \begin{center}
    \subfigure[]{
      \includegraphics[width=0.28\textwidth]{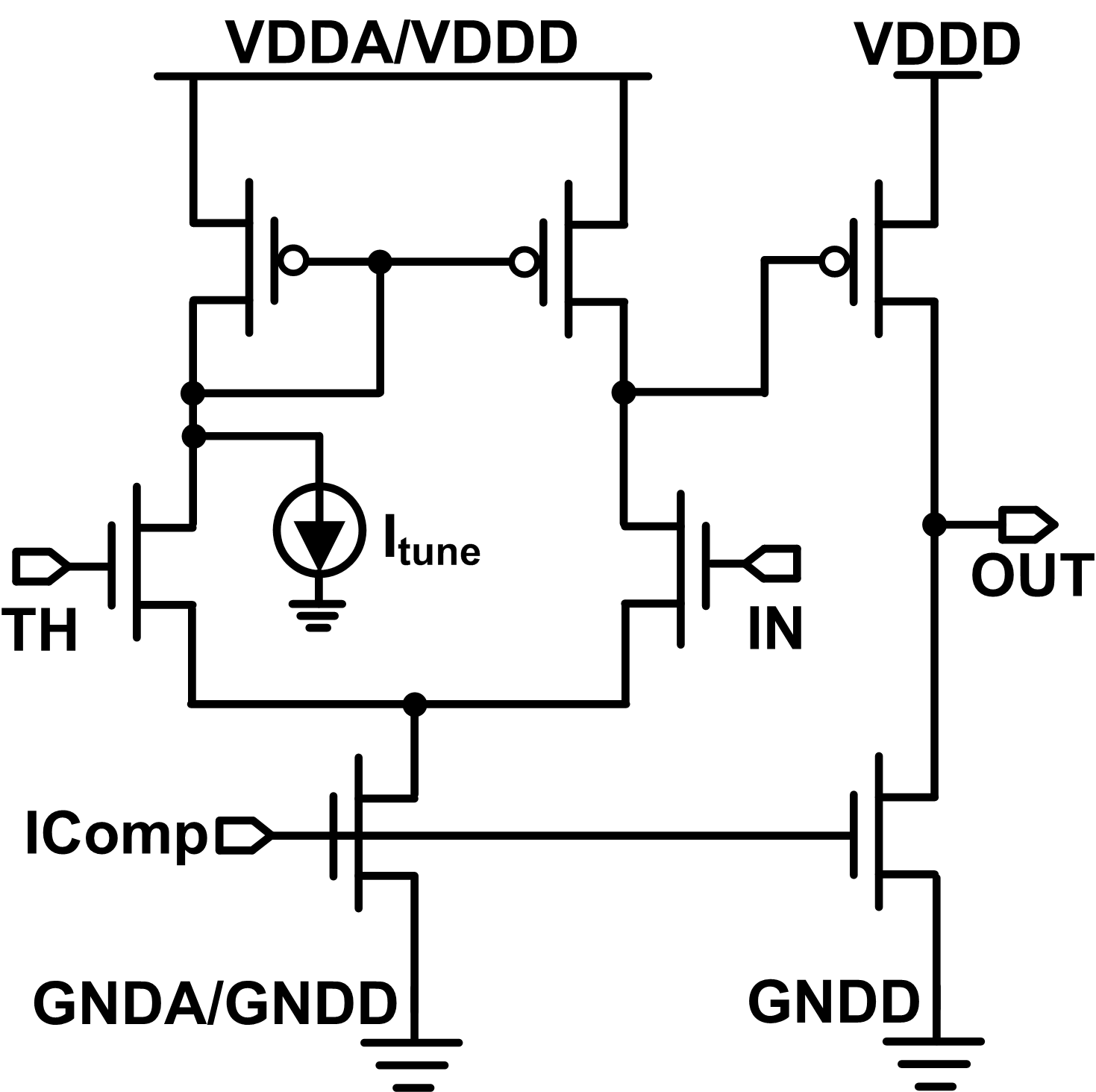}
      \label{fig:DisV1}
    }
    \qquad
    \qquad
    \subfigure[]{
      \includegraphics[width=0.33\textwidth]{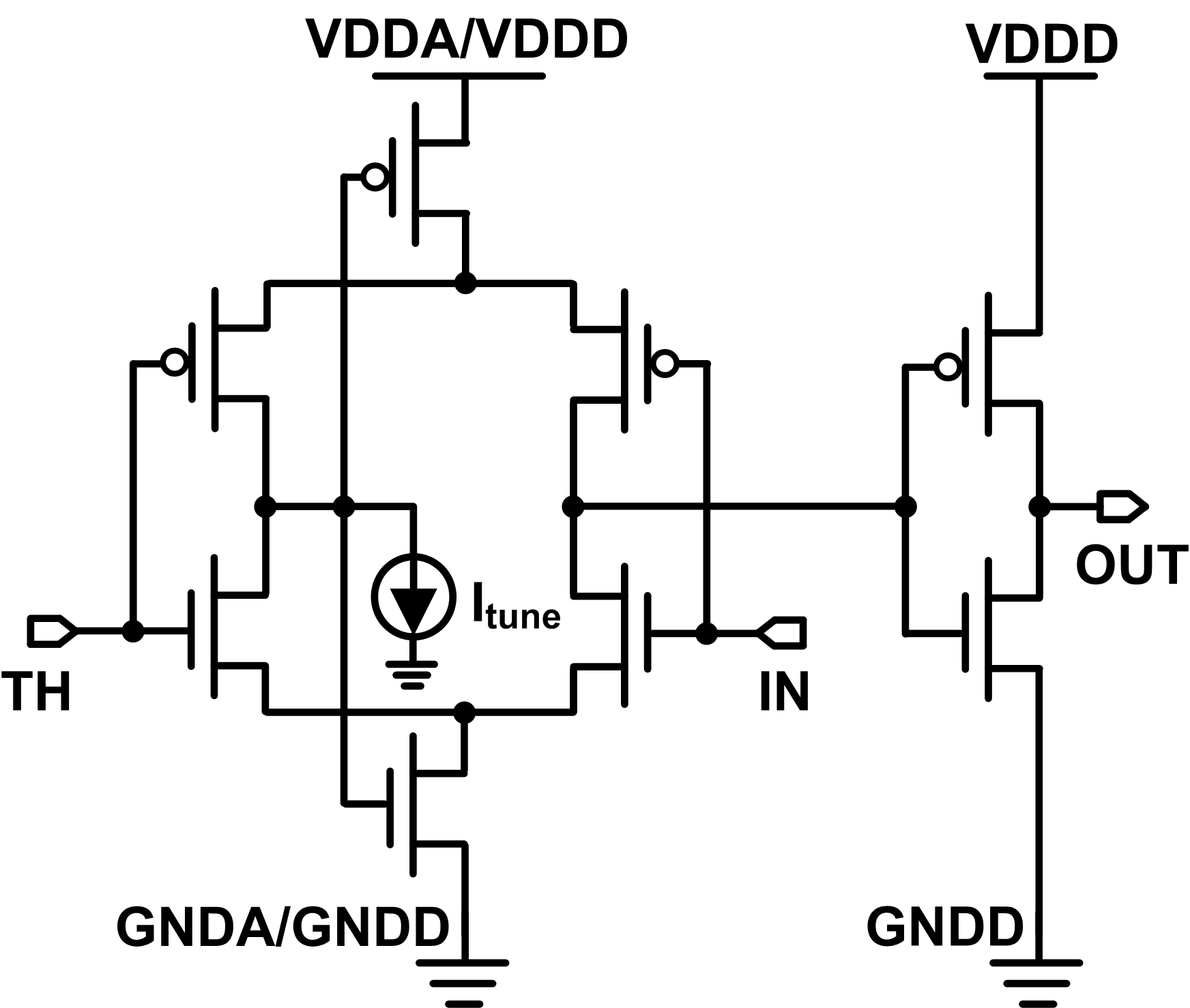}
      \label{fig:DisV2}
    }
    \caption[]{Schematic of \subref{fig:DisV1} discriminator V1 and \subref{fig:DisV2} discriminator V2.} 
    \label{fig:Dis}
  \end{center}
\end{figure}

\paragraph{Discriminator} Figure\,\ref{fig:Dis} shows the schematics of the two discriminator designs. The discriminator V1 in figure~\ref{fig:DisV1} employs a typical two-stage open loop amplifier, the same circuit as used in \mbox{LF-CPIX}. The discriminator V2 in figure~\ref{fig:DisV2} is new in LF-Monopix1, and it features a self-biased differential amplifier~\cite{Bazes_1991} followed by a CMOS inverter as the output stage. The latter exploits the complementary characteristics of CMOS input devices for high gain and fast operation. The nominal current consumption is 4.5\,$\upmu$A for discriminator V1 and $\sim$3.5\,$\upmu$A for discriminator V2. The pixel-to-pixel threshold dispersion can be mitigated out using a threshold tuning current I$_{\text{tune}}$ per pixel, which can be adjusted by an in-pixel 4-bit current DAC. The discriminator is the interface between the analog and the digital domain. Two powering schemes are implemented in LF-Monopix1 to examine the possible interference between power domains, i.e.\ either using the digital power supply only or using the analog power supply for the first stage.

\subsubsection{In-pixel digital readout circuit}
\label{sec:PixelLogic}

\begin{figure}[htbp]
  \centering
  \includegraphics[width=.95\textwidth]{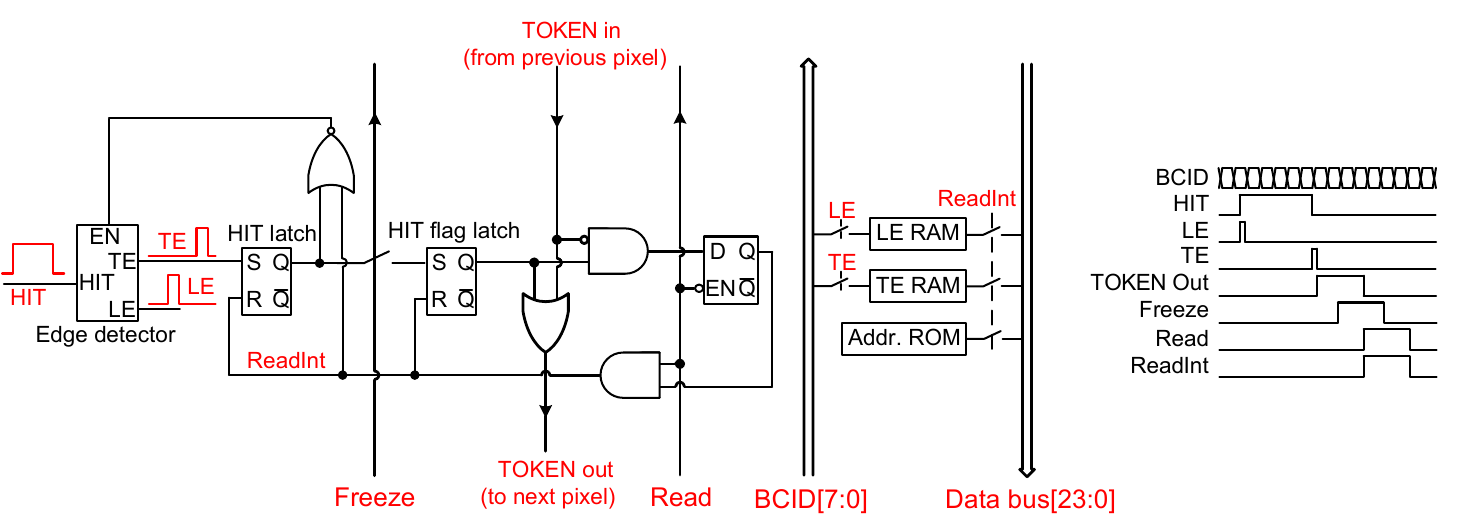}
  \caption{Schematic of the in-pixel readout logic. The timing diagram for operation is shown on the right.}
  \label{fig:PixelRO}
\end{figure}

Figure\,\ref{fig:PixelRO} shows the functional view of the in-pixel digital circuit implemented in LF-Monopix1, together with its operation timing diagram. The circuit receives a digital pulse \emph{HIT} from the discriminator. An edge detector detects the leading edge \emph{LE} and trailing edge \emph{TE} of the \emph{HIT} signal, generating two short pulses ($\sim$\,1\,ns) that strobe their corresponding time stamps into the in-pixel RAM cells. The \emph{LE} time stamp gives the Time of Arrival (ToA) for the hit, and the difference between \emph{TE} and \emph{LE} time stamps provides analog information in terms of Time over Threshold (ToT). The 8-bit timing information is provided by a timing bus, distributing differentially the 40\,MHz gray-encoded beam Bunch Crossing ID (\emph{BCID}) along the pixel column. The gray code minimizes the digital switching, and accordingly the power consumption and digital cross talk. Additionally, differential lines further mitigate any cross talk to the neighboring signal lines. 

The readout of pixel hits is arbitrated by a token passing scheme. The \emph{TE} pulse sets the \emph{HIT flag} latch, which asserts a \emph{TOKEN} signal that propagates to the column end and signals the appearance of hits in the column. Upon receiving the \emph{TOKEN}, a readout controller (not shown in figure~\ref{fig:PixelRO}) sends back a \emph{Freeze} signal to the column, that prohibits the \emph{HIT flag} latch from being set by later hits. This makes sure that any new hit will not disturb the readout process, but can still be recorded by the first \emph{HIT latch}. The readout controller then generates a \emph{Read} signal, passing by a priority network in the column. The result is that only the hit pixel with the highest readout priority, i.e.~the hit pixel without a valid \emph{TOKEN} input signal, has its internal \emph{ReadInt} signal activated. The \emph{ReadInt} signal allows the hit information (8-bit \emph{LE} time stamp, 8-bit \emph{TE} time stamp and 8-bit pixel address) from the in-pixel memories to be driven to the column data bus for readout. Since the \emph{ReadInt} signal also clears the \emph{HIT flag} register, the priority network will ripple down to find the next hit pixel, if any, once the \emph{Read} signal goes back to zero. This readout cycle continues by sending repetitively the \emph{Read} pulse until this frozen column is "drained out".

The in-pixel digital circuit employs full custom design in order to minimize its area, and accordingly the detector capacitance contribution from C$_{\text{pw}}$ (see section~\ref{sec:lfdev_challenges}). Special care has been taken to mitigate the digital switching noise. For example, a current steering logic \cite{Hiok_1997} is used to propagate the token signal, and source followers are used as the line driver for the column data bus in order to avoid excessive current spikes during data readout.

The pixel digital processing logic can also be placed at the chip periphery, in which case one-to-one connection from the pixel AFE output to its corresponding logic unit at the periphery is needed. This scheme, adopted also in\,\cite{Peric_2013}, minimizes the amount of in-pixel electronics, thus reducing the PSUB/PW area and the corresponding C$_{\text{pw}}$. The absence of digital in-pixel circuit also makes the design less prone to cross talk. 
However, as a penalty, the routing resources from the pixels to the periphery are very demanding and the insensitive periphery area is increased, especially when scaling up the pixel array to a large size. For comparative studies, pixel designs with logic at the periphery have also been implemented in LF-Monopix1.

\subsubsection{Layout implementation}
\label{sec:Layout}
\begin{figure}
  \centering
  \includegraphics[width=.9\textwidth]{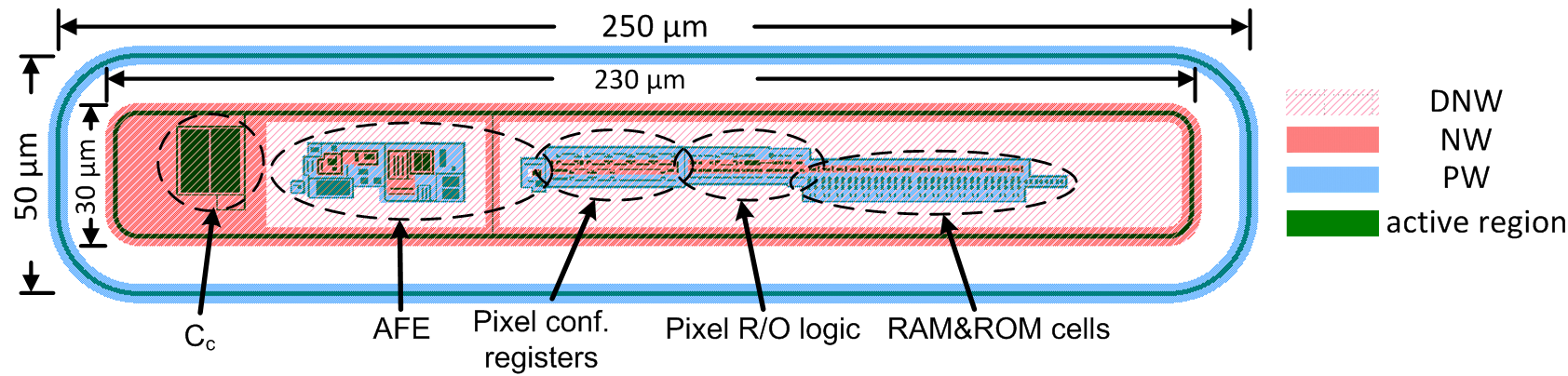}
  \caption{Layout of a typical pixel design. The red shaded area represents the DNW implant. The solid red area is the NW implant above DNW. The NW ring structure also has a buried NISO layer joining it to DNW (refer to figure\,\ref{fig:crosssection}).  NWs and PWs inside the DNW are used to host transistors. The NWs for PMOS transistors (except for the AC coupling MOS capacitor $\mathrm{C_{c}}$) need to be isolated from the DNW, as these NWs are connected to the circuit power supply instead of the sensor bias potential and will introduce parasitic charge collection paths. Therefore, these NWs are surrounded by PWs, and a PSUB layer is required below them (refer to figure\,\ref{fig:crosssection}). The outer PW ring is the p-stop implant which provides isolation from neighboring pixels. The active region is typically used for implementing active devices or generating heavily doped contact regions on the silicon surface.}
  \label{fig:PixLayout}
\end{figure}

The layout implementation of a typical pixel cell is depicted in figure~\ref{fig:PixLayout}. The charge collection DNW takes about 55\,\% of the total pixel area. The PW ring surrounding the DNW works as a p-stop implant which stops the conducting channel arising from electron  accumulation at the interface between silicon and silicon dioxide. In order to minimize the interference from the digital circuit to the sensitive analog region, the digital circuit is located in a separate ``bulk'' (PSUB and PW) and operates in its own power domain. It is noted that the potential of the digital ``bulk'' needs to be kept as stable as possible to minimize the cross talk to the sensing node via C$_{\text{pw}}$. Therefore, the digital ``bulk'' is not connected to the digital ground which tends to be noisy, and is tied to the ground potential with a dedicated metal line.

\subsection{Chip architecture}
\label{sec:chip}
\begin{figure}
  \centering
  \includegraphics[width=.9\textwidth]{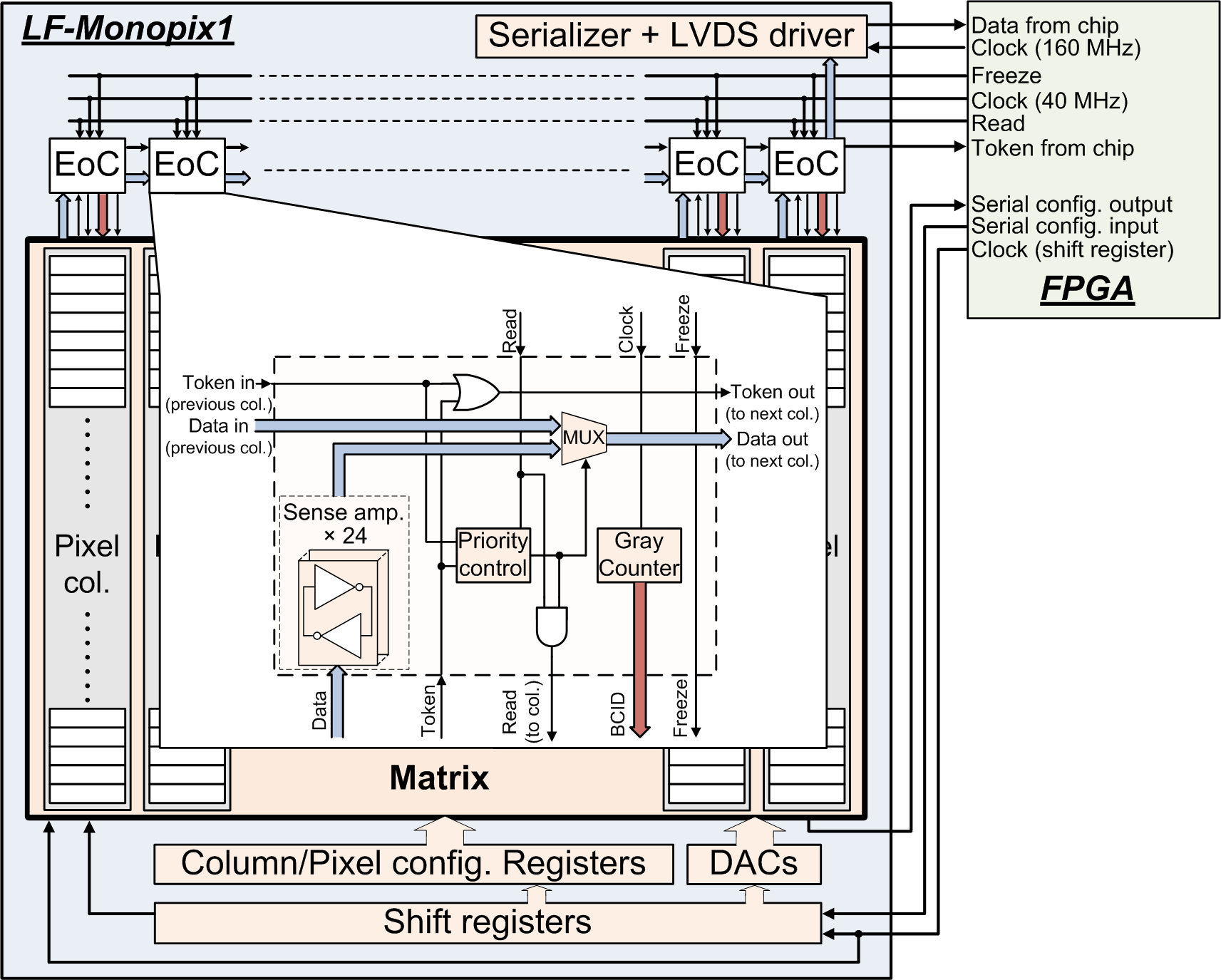}
  \caption{Block diagram of the LF-Monopix1 architecture. Explanations are given in the text.}
  \label{fig:Chip}
\end{figure}

The block diagram for the LF-Monopix1 chip is shown in figure\,\ref{fig:Chip}. There are nine different pixel designs implemented in an array of 129\,$\times$\,36 pixels, with each design variant taking four columns. Table\,\ref{tab:Matrix} shows the distribution of different pixel designs in the matrix. Seven designs are pixels with built-in readout logic. They differ from each other by different circuit designs and layout implementations (see section\,\ref{sec:pixel}). The remaining two designs have their pixel digital processing logic located at the column-end periphery.

\begin{table}
    \centering
    \caption{Distribution of pixel variants in the matrix}
    \label{tab:Matrix}
    \smallskip
    \begin{minipage}{\textwidth}
        \renewcommand\footnoterule{}
        \begin{tabular}[t]{l|l|l|l|l|l|l|l|l|l}
            \hline
            Column & 0-3 & 4-7 & 8-11 & 12-15 & 16-19 & 20-23 & 24-27 & 28-31 & 32-35 \\
            \hline
            Pixel R/O logic & \multicolumn{2}{c|}{out pixel} & \multicolumn{7}{c}{in pixel} \\
            \hline
            Preamplifier & \multicolumn{7}{c|}{CMOS} & \multicolumn{2}{c}{NMOS} \\
            \hline
            Discriminator & \multicolumn{2}{c|}{V2} & V1 & V2 & V1 & V2 & V1 & V2 & V1 \\
            \hline
            Discri. power\footnote{D: Digital~~~~~~~~~~~~~~~~~~~~~~~~~~~~~~~A: Analog\label{power_domain}} & \multicolumn{2}{c|}{D} & A+D & \multicolumn{2}{c|}{D} & \multicolumn{4}{c}{A+D} \\
            \hline
            Token logic & \multicolumn{3}{c|}{CMOS logic} & \multicolumn{6}{c}{Current steering logic} \\
            \hline
            Col. line driver\footnote{P: PMOS source follower~~~~~~N: NMOS source follower.\label{line_driver}} & P & \multicolumn{8}{c}{N} \\   
            \hline
        \end{tabular}
    \end{minipage}
\end{table}

The bottom part of the prototype includes the circuits for chip bias and configuration. The chip is configured through a chain of shift registers located both at the chip periphery and inside the matrix. There are also dedicated monitoring lines that can broadcast the CSA and the discriminator outputs of a selected pixel off the chip for debugging and chip characterization. On the top side, each pixel column interfaces with its own End-of-Column (EoC) circuitry. The EoC circuitry includes 24 sense amplifiers to receive and temporally store the 24-bit hit information from the column data bus, a gray counter running at 40~MHz to provide the \emph{BCID}, and the column readout logic that performs the column level priority scan and data transmission. The data output from the EoC circuit is serialized and sent out by an LVDS driver with a bit rate of 160 Mbps. As the goal of LF-Monopix1 is to demonstrate a large column drain readout matrix, the readout controller, which controls chip configuration and readout sequence, is implemented off chip in an FPGA in order to simplify the chip design. There are no data buffering memories at the periphery. Triggered readout as required for the LHC pp-experiments is not implemented either. This means that all the hit pixels in the matrix are read out sequentially through the output serial link following a pre-defined priority.

\section{Characterization of the prototypes}
\label{sec:res}
The goal of the development is a monolithic depleted pixel sensor, capable to sustain the radiation levels encountered in a pixel tracker in an HL-LHC-type radiation environment. 
In the outer pixel layers of the ATLAS Inner Tracker (ITk), for example, the radiation exposure amounts to a total ionization dose (TID) of 50\,Mrad and to a non-ionizing energy loss (NIEL) particle fluence with a damage equivalent to $10^{15}$ of 1-MeV neutrons per cm$^2$. 
The chips were irradiated with protons, neutrons or X-rays. 
Proton and neutron irradiated chips are mainly used to characterize radiation damage of the sensing part in the DMAPS chips.
As charged particles, protons also cause ionization damage, which mainly affects the electronics. X-ray irradiation is used to characterize 
ionizing damage to the electronics part of the chips.

Proton irradiation was carried out at the Proton Irradiation Facility at CERN \cite{irrad} with calibrated NIEL and TID damages corresponding to the above specified target values, which are TID of 50\,Mrad and fluence of $10^{15}~\mathrm{n_{eq}/cm^2}$ .
Neutron irradiations were performed at the TRIGA Mark II Research Reactor in the Jo\v{z}ef Stefan Institute \cite{jsi} to fluences of $10^{14}~\mathrm{n_{eq}/cm^2}$ and $10^{15}~\mathrm{n_{eq}/cm^2}$.
The irradiated chips were annealed for 80\,min at 60{$^\circ$}C.
X-ray irradiation was performed using a 2\,kW tungsten X-ray tube with 100\,kV acceleration voltage.
The Irradiation Center at the Karlsruhe Institute for Technology \cite{kit} provided the X-ray radiations.
The top side of the chips, the side which contains the electronics, was illuminated with X-rays to a dose up to 50 Mrad with a rate of 572~krad/h.
The irradiation was paused 22~times and measurements were performed after each TID step.

\subsection{Biasing of sensor diode}
\begin{figure}
    \centering 
    \includegraphics[width=0.6\linewidth,trim=9.0cm 15.0cm 0.0 1.0cm,clip]{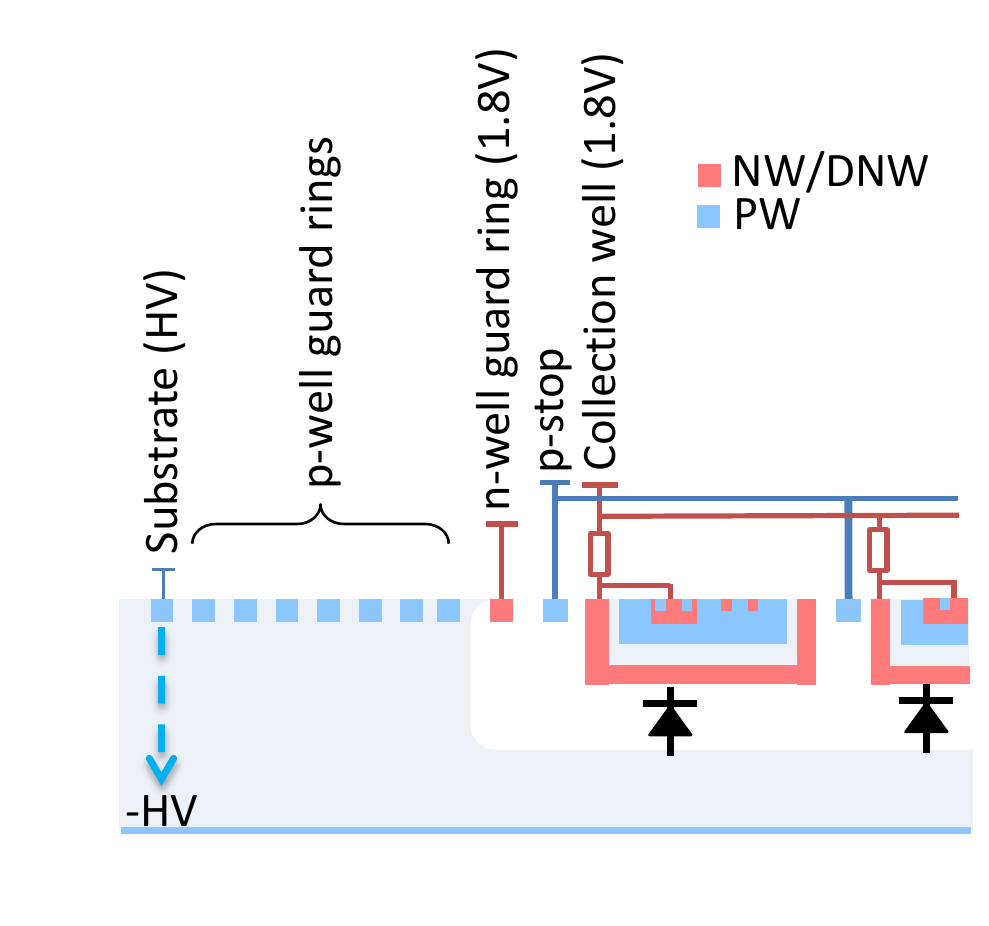}
    \caption{\label{fig:guardrings} Cross-sectional view of LF-Monopix1 near the chip edge. The color coding is indicated in the figure. There are eight p-well implant rings and one n-well implant ring surrounding the pixel matrix and the periphery.
    }
\end{figure}

%
%
The three prototype chips discussed in this paper possess similar guard ring structures surrounding the pixel matrix and the periphery.
Figure \ref{fig:guardrings} shows the cross sectional views of the LF-Monopix1 chip.
Eight p-well implant rings and one n-well implant ring are used.
The outer most p-well ring acts as a biasing node to which negative HV is applied. 
The remaining seven p-rings together with the n-ring act as guard rings protecting the matrix e.g. against leakage currents.
The n-well ring is biased at a voltage of 1.8~V while all other p-well rings as well as the p-well structure between pixels are electrically floating. 
Each collection node is connected to a biasing circuit inside the pixel, resulting in a potential of approximately 1.8\,V.

\begin{figure}
    \centering 
    \includegraphics[trim=0cm 0 0 0,clip]{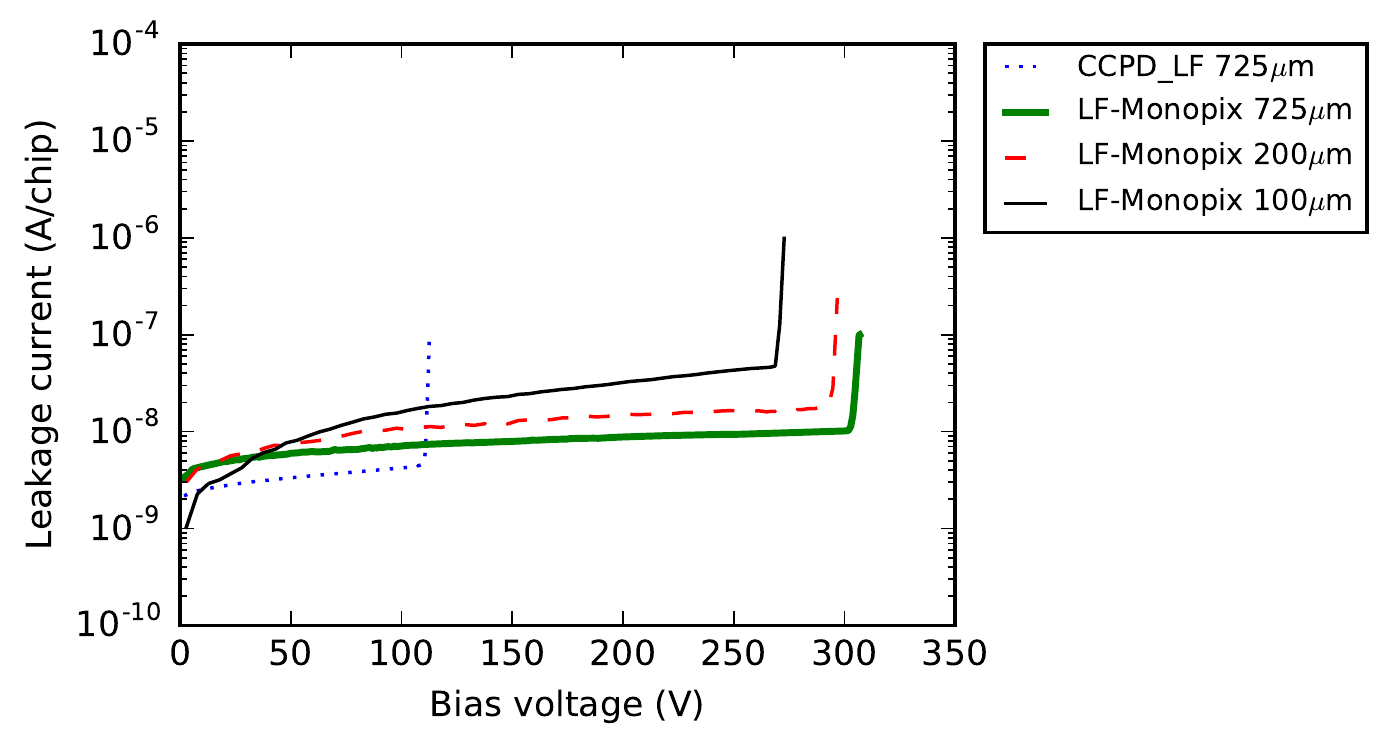}
    \caption{\label{fig:iv1} I-V curves of CCPD\_LF (dotted line) and LF-Monopix1 chips (other lines).
    The various chip thicknesses are given in the legend.
    The breakdown voltage of CCPD\_LF is 115~V. 
    For LF-Monopix1 chips, the breakdown voltages are 270\,V, 295\,V and 304V for 100\,$\upmu$m, 200\,$\upmu$m, and 750\,$\upmu$m thickness, respectively.
    }
\end{figure}

Figure \ref{fig:iv1} shows the current-to-voltage (I-V) curves of the prototype chips.
The breakdown voltage of the first prototype (CCPD\_LF) is 115\,V. 
The location of the breakdown point was determined using a photon-emission microscope and located between the innermost p-well ring (floating) and the n-well ring (on 1.8~V) \cite{Hirono_Thesis}.
For the second and third prototype the guard-ring structure was optimized by increasing the distance between the innermost p-well ring and n-well ring, resulting in a much higher breakdown voltage. 
As is evident from figure\,\ref{fig:iv1} the breakdown voltage exceeds 300\,V and is still higher than 250\,V for sensors thinned down to 100\,$\upmu$m.

The thickness of the high-resistive substrate wafer is 725\,$\upmu$m. 
Thinning and back-side processing were applied to obtain the desired sensor thicknesses.
The wafers were ground to 100\,$\upmu$m and 200\,$\upmu$m using the TAIKO process \cite{taiko}. 
Then, plasma etching, boron implantation, and annealing were applied to the wafers.
For the 200\,$\upmu$m thick wafers, metallization was added after the implantation.
Since the three measured LF-Monopix1 chips originated from different wafers, wafer variants and chip variants could also explain the observed difference of the breakdown voltages. 
Nevertheless, we can conclude that the design features a very high breakdown voltage ($\gtrsim$\,200\,V) even after thinning and back-side processing.

\begin{figure}
    \centering 
    \includegraphics[width=\textwidth,trim=0cm 0 0 0,clip]{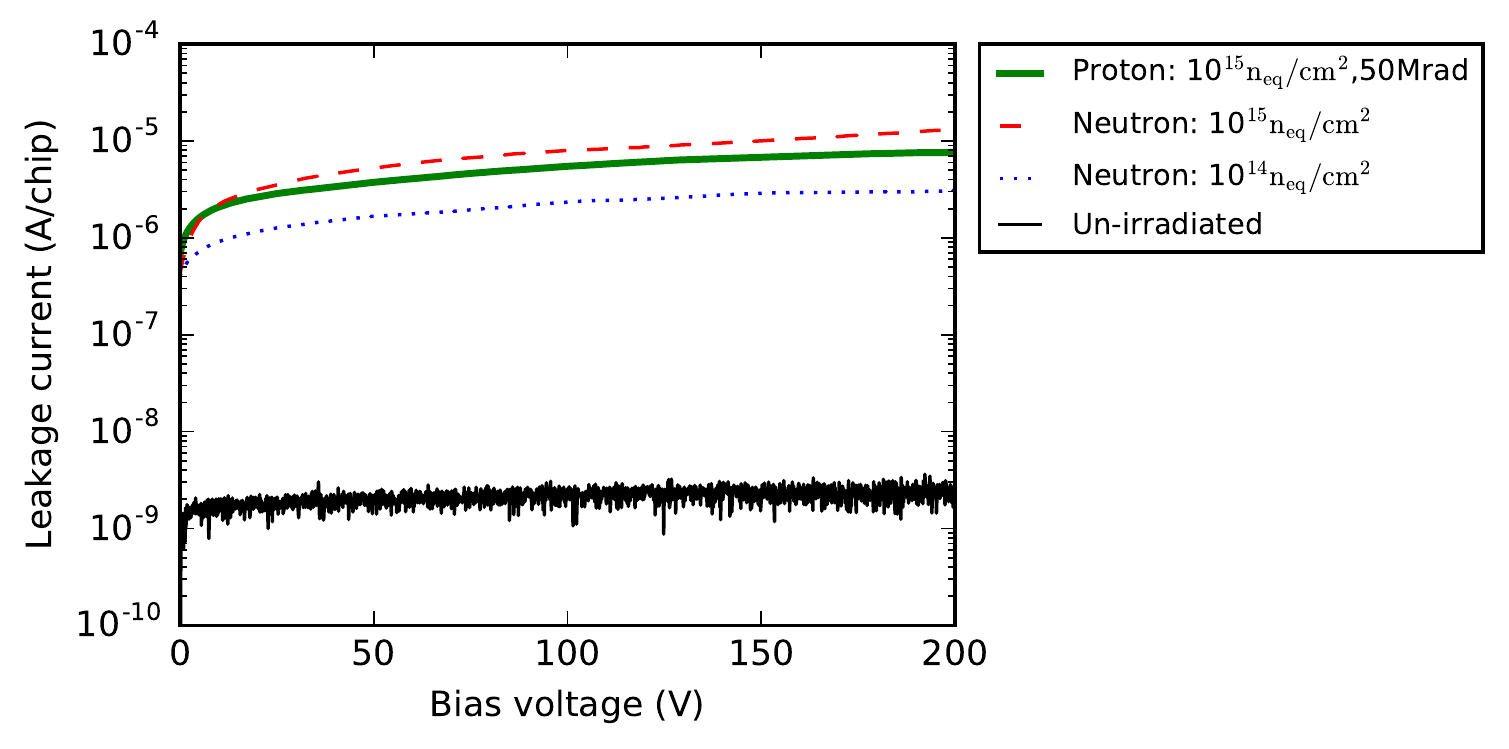}
    \caption{\label{fig:ivProtonLFMonopix200} I-V curves of neutron and proton irradiated LF-Monopix1 chips.
    Type of particle used in irradiation and its fluence are shown in the legend.
    The un-irradiated performance is also shown for a comparison. 
    The measurements were performed in a climate chamber at an atmospheric temperature of -25$^\circ$C.}
\end{figure}
The I-V curves of LF-Monopix1 prototypes were also measured after proton or neutron irradiation, respectively.
The chips were cooled during the measurement and their leakage current was kept below 0.1\,mA in order to protect the chips from self-heating thermal runaway. 
Figure \ref{fig:ivProtonLFMonopix200} shows the I-V curves up to bias voltages of 200\,V. 
Higher voltages were not applied to prevent the chips to be broken by applying unexpected large current.
The measurement results show that breakdown occurs well above 200 V.

\subsection{Radiation hardness of the analog front-end circuit}
The biggest concern regarding the amplifier stages of the three CSA types realised in LF-CPIX and LF-Monopix1 is their radiation hardness against ionizing radiation.  Because the CSA designs realized in LF-Monopix1 (see section\,\ref{sec:FE}) are largely adopted from those implemented in LF-CPIX, the results obtained on the CSA circuits in LF-CPIX are representative for both chips. LF-CPIX chips have been irradiated by X-rays to evaluate the TID hardness, in particular of its analog front-end circuitry.

The three CSA design flavours implemented in LF-CPIX feature different types of input devices (see figure\,\ref{fig:PreAmp}): an NMOS transistor, a PMOS transistor or two complementary transistors (CMOS). The NMOS and CMOS CSAs are almost identical to those in LF-Monopix1, while the PMOS CSA in LF-CPIX is an improved version of that in CCPD\_LF \cite{Hirono_2016_IEEE}.

\begin{figure}[h]
    \centering 
    \includegraphics[trim=0 0 0 0,clip]{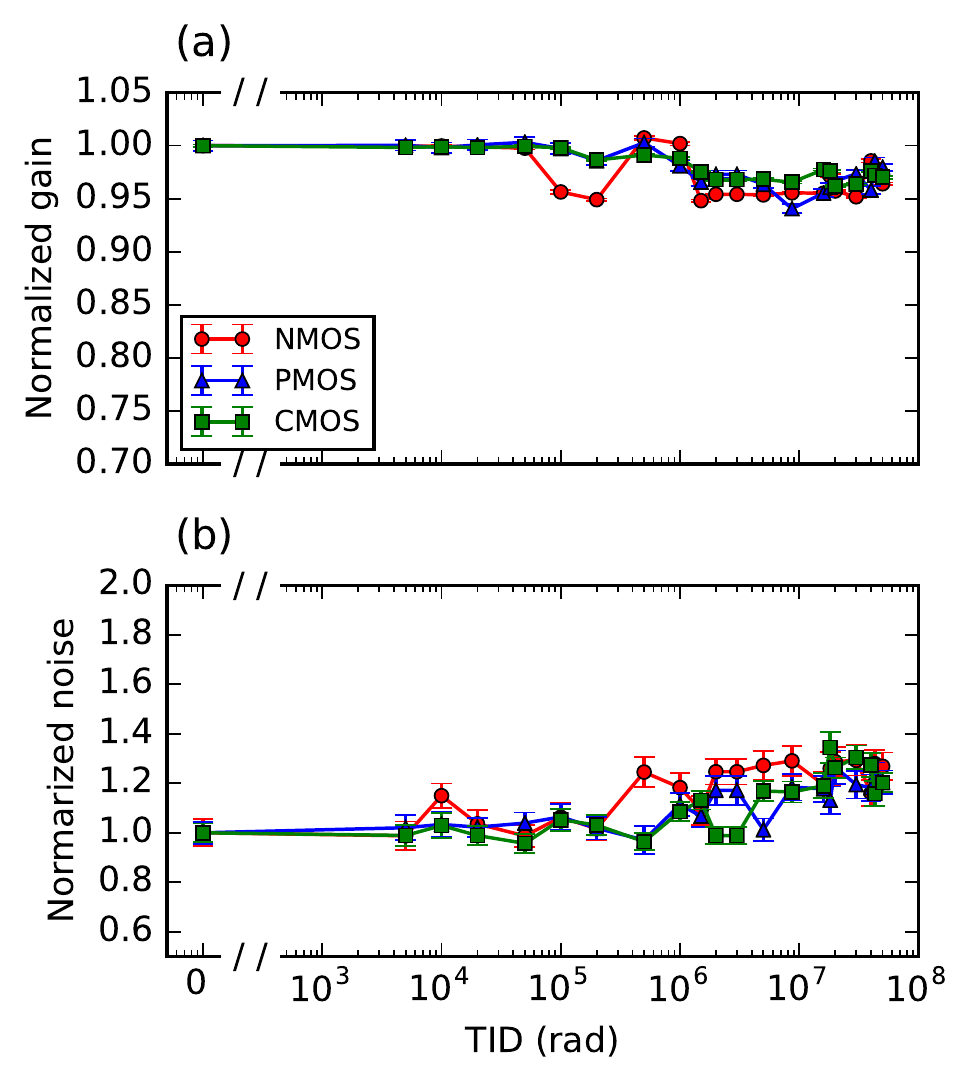}
    \caption{\label{fig:TIDgain} (a) Normalized gain and (b) ENC of LF-CPIX pixels.
    Pixels with NMOS, PMOS, and CMOS CSA are irradiated up to 50~Mrad using X-ray tube.
    The gain and ENC are normalized to values before irradiation. The error bars are statistical  only from fits to charge injections scans.}
\end{figure}

Figure \ref{fig:TIDgain} shows the normalized gain and equivalent noise charge (ENC) of the LF-CPIX chip throughout the X-ray irradiation up to 50\,Mrad. 
The shown errors are statistical only from fits to charge injection scans. The systematic
uncertainties and fluctuations are much larger than the statistical errors. 
The trend of the measurements is, however, clear in both figure parts.
The three CSA designs behave very similarly, having a gain degradation of $\sim$\,5\% and a noise increase of less than 30\%. 
The noise increase is explained by the increase of the leakage current with radiation. 
The measured leakage current is 3\,$\upmu$A/chip with an applied bias voltage of 200\,V at room temperature. It is expected to decrease by cooling the chip to typical LHC operation temperatures of 
-20$^\circ$C.

\begin{figure}
    \centering 
    \includegraphics[trim=0 0 0 0,clip]{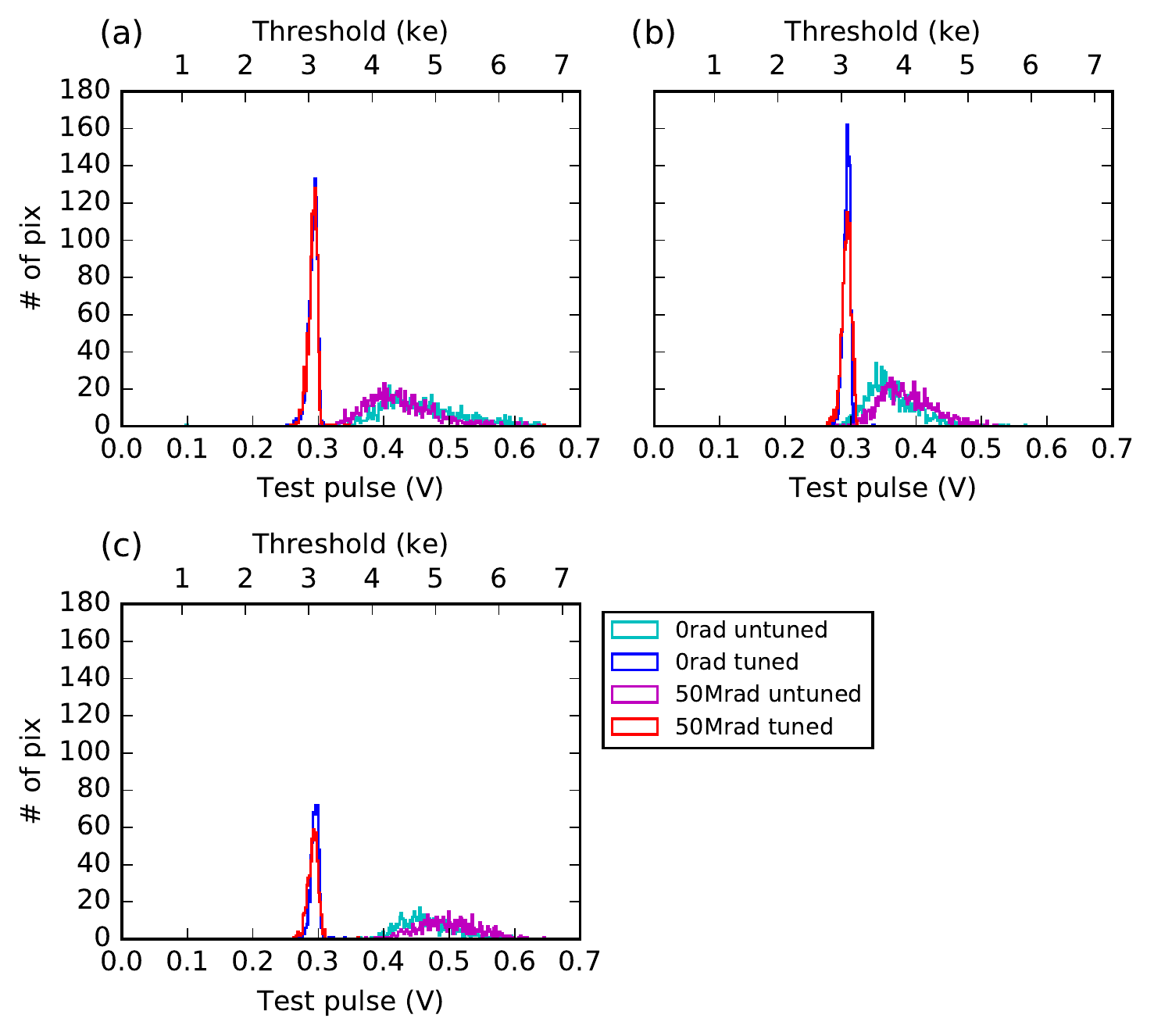}
    \caption{\label{fig:TIDth} Distributions of pixel discriminator thresholds before and after X-ray irrdiation to doses of 50 Mrad: (a) NMOS, (b) PMOS and (c) CMOS CSA. The color coding is indicated by the box. Note that the number of pixels entering (c) is roughly 45\% lower.
     }
\end{figure}

Each pixel has a 4-bit trim DAC to tune the threshold voltage as described in section\,\ref{sec:FE}.  
Figure \ref{fig:TIDth} shows the threshold distributions before and after X-ray irradiation, with and without applying threshold tuning.
The distributions are fitted with Gaussians. The spreads ($\sigma$) before irradiation are 67.8\,e$^{-}$, 54.3\,e$^{-}$ and 63.0\,e$^{-}$ for NMOS, PMOS and CMOS CSAs, respectively.
The spread is enlarged after TID irradiation but the threshold can still be tuned to a sharply peaking distribution. 
The spreads after irradiation are 64.5\,e$^{-}$, 76.2\,e$^{-}$ and 83.5\,e$^{-}$ for each CSA type with typical errors in the order of 5-10\,e$^{-}$.
The spread of the un-tuned distributions increases negligibly for the CMOS and the PMOS CSAs to approximately 20\,e$^{-}$. There is no increase in spread observed for the NMOS CSA.
The threshold dispersions of both LF-CPIX and LF-Monopix1 have additionally been investigated in a proton irradiation campaign up to a level of 150\,Mrad \cite{Chen_2019}.

\begin{figure}
    \centering 
    \includegraphics[trim=0 0 0 0,clip]{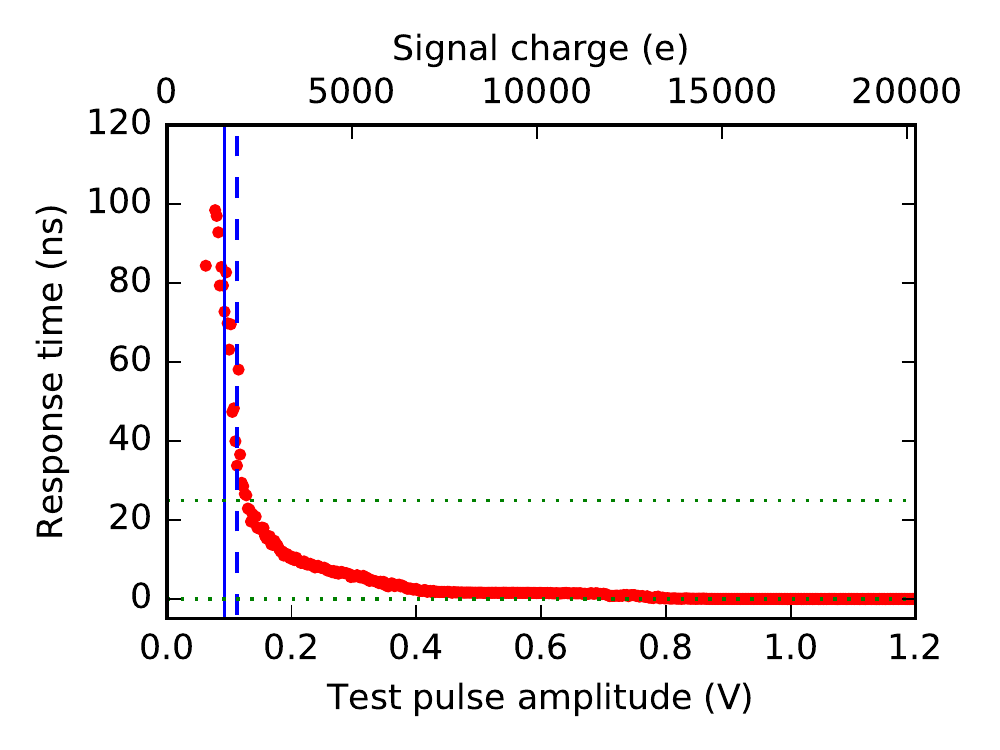}
    \caption{\label{fig:tw} Relative response time of a pixel with NMOS CSA and V2 discriminator in LF-Monopix1 as a function of the signal amplitude in volts (bottom axis) and in corresponding electron charge (top x-axis).
    The offset of the response time is calibrated to zero for very large signals (test pulses of 1.2\,V).
    The dotted lines indicate the range of signal pulses responding within 25\,ns.
    The discriminator threshold at 1554\,e$^{-}$ (solid line) and the in-time threshold at 1896\,e$^{-}$ (dashed line) are also shown. 
    The ``overdrive'' is 342\,e$^{-}$.}
\end{figure}

The timing performance of LF-Monopix1 is measured using test pulse injection for a threshold setting of 1550\,e$^{-}$.
Figure \ref{fig:tw} shows the response time of a discriminator as a function of the test pulse amplitude for a pixel with NMOS CSA and V2 discriminator.
The circuit response to larger signals is faster than to smaller signals.
The response time to a very large signal amplitude (here at 1.2\,V, corresponding to 20.2\,ke$^{-}$) is therefore chosen as calibration zero and the response delay relative to this large signal is plotted. 
Since the HL-LHC collision frequency is 40~MHz, the variance of response time in one collision must be smaller than 25~ns.
The response of the discriminator can still be registered within the required LHC bunch crossing window of 25\,ns for signals larger than 1896\,e$^{-}$ (in-time threshold = an effective threshold). The ``overdrive'' is the difference between the threshold and the in-time threshold (here 340\,e$^{-}$).

\begin{figure}
    \centering 
    \includegraphics[trim=0 0 0 0,clip]{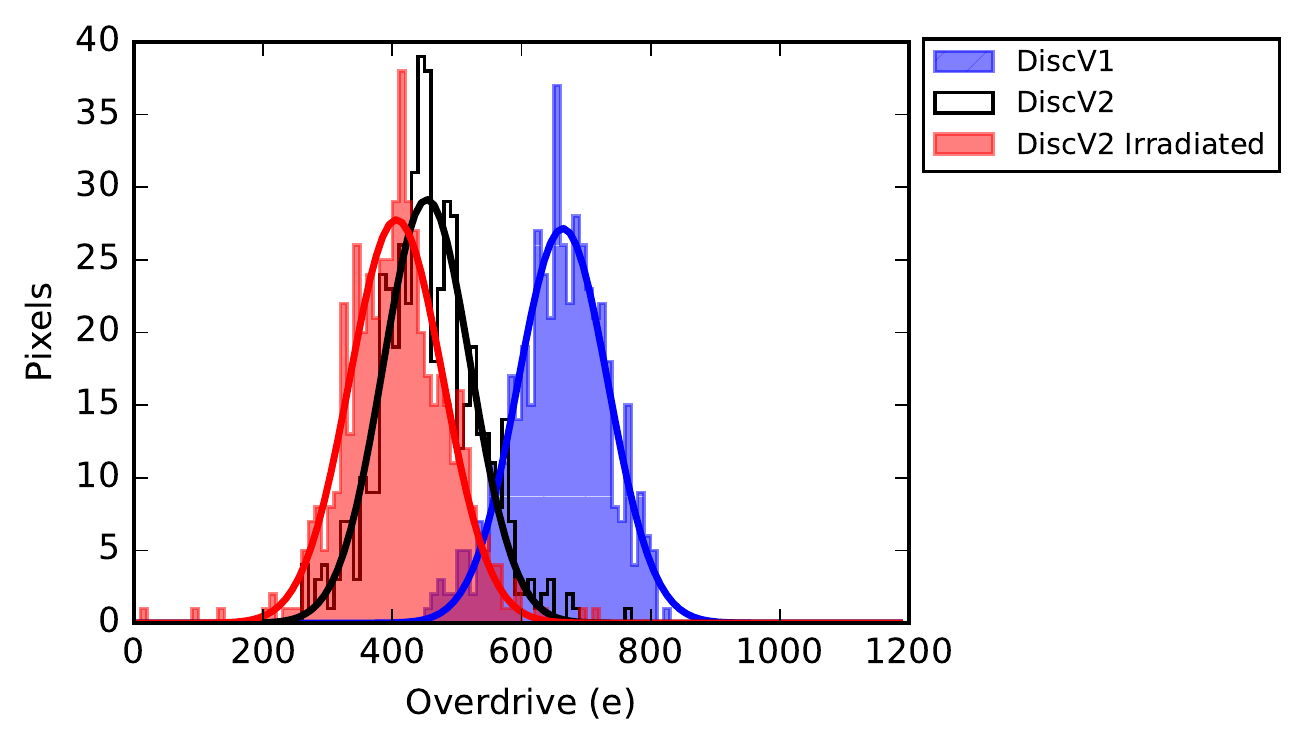}
    \caption{\label{fig:pixtw} ``Overdrive'' distributions of un-irradiated as well as for proton-irradiated ($10^{15}\mathrm{n_{eq}/cm^2}$, 50~Mrad) LF-Monopix1.
    For the un-irradiated chip two types of discriminator are shown separately.
    The threshold of each pixel is set to 1500\,e$^{-}$.
    Solid lines represent fitted Gaussians. 
    The fitted overdrive for un-irradiated V1, un-irradiated V2 and irradiated V2 are 664~e$^{-}$, 453~e and 406~e$^{-}$, respectively.}
\end{figure}

The ``overdrive'' distributions of LF-Monopix1 are shown in figure \ref{fig:pixtw} for two different discriminator designs as
detailed in section \ref{sec:design}. The operating parameters of the CSA and the discriminators were optimized to minimize the time walk.
The average ``overdrives'' are 664\,e$^{-}$ and 453\,e$^{-}$ for V1 and V2, respectively.
The pixels with V2 discriminator show a smaller overdrive than V1 discriminator as expected from design simulations.

The overdrive of the proton-irradiated (and annealed) chip is also included in figure \ref{fig:pixtw}.
Mean value and sigma of the overdrive determined by a Gaussian fit are 406\,e$^{-}$ and 72\,e$^{-}$, respectively. The mean shows only a small shift to lower values compared to the un-irradiated sample, which could well be due to chip-to-chip variations, for a precise
assessment of which statistics is not sufficient.
Nevertheless, no serious degradation in timing performance observed due to TID damage is concluded.

\subsection{Efficiency before and after irradiation}
\begin{figure}
    \centering 
    \includegraphics[trim=0 0 0 0,clip]{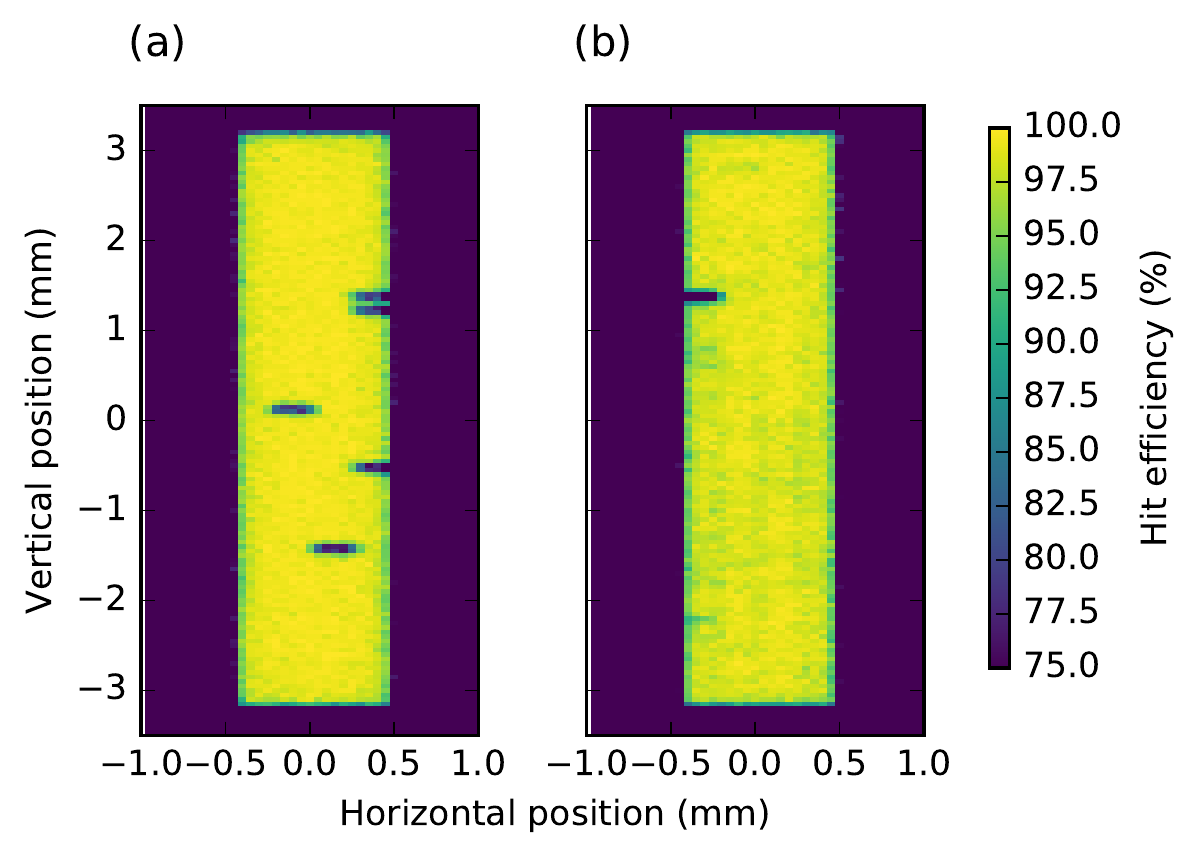}
    \caption{\label{fig:Eff} Hit detection efficiency map of (a) un-irradiated and (b) neutron irradiated ($10^{15}\mathrm{n_{eq}/cm^2}$) LF-Monopix1.
        The hit detection efficiencies are (99.7$\pm$0.1)\% and (98.9$\pm$0.1)\% for un-irradiated and irradiated chip, respectively. Non-efficient areas correspond to masked pixels. Five and one pixel were masked in the un-irradiated and irradiated chip, respectively. The masked regions and edges of measured area were excluded from the efficiency calculation. }
\end{figure}

The hit detection efficiency of LF-Monopix1 before irradiation and after neutron irradiation has been measured in a dedicated test beam using the 2.5\,GeV electron beam at the External Beamline for Detector Tests in the Electron Stretcher Accelerator at the University of Bonn \cite{elsa:e3xd}.
The beam track positions on the LF-Monopix1 chips were obtained using ANEMONE, an EUDET-type beam telescope \cite{eudet} operated by a newly developed DAQ system \cite{pascal:thesis}.
The threshold was set to 1800\,e${^-}$ and 1600\,e${^-}$ for the un-irradiated and the neutron irradiated chip, respectively.
At these thresholds, the noise occupancy of the chips is more than an order of magnitude smaller than the HL-LHC requirement of 10$^{-6}$ hits per pixel and bunch-crossing time (25~ns), corresponding to a hit rate of 40\,Hz/pixel. The chips were cooled below -40$^\circ$C by dry ice.
The bias voltage applied to the chips was 200\,V. The resulting hit detection efficiency is shown in figure\,\ref{fig:Eff}.
The dark blue spots corresponding to low hit detection efficiency are disabled (masked) pixels which could not be tuned to the target thresholds and are excluded from the efficiency calculation.
The telescope resolution of the reconstructed tracks is about 20\,$\upmu$m. Each hit in LF-Monopix1 is assigned to a reconstructed track if it is found within a distance of 300\,$\upmu$m for column- and row-wise direction of the LF-Monopix1 matrix.
The average hit detection efficiencies, where pixels are enabled, are (99.7$\pm$0.1)\% and (98.9$\pm$0.1)\% for the un-irradiated and the irradiated chip, respectively. 
The dominant error of this measurement is due to miss-assignment between the beam track extrapolation and a given hit in LF-Monopix1. 

\begin{figure}
    \centering 
    \includegraphics[trim=0 0 0 0,clip]{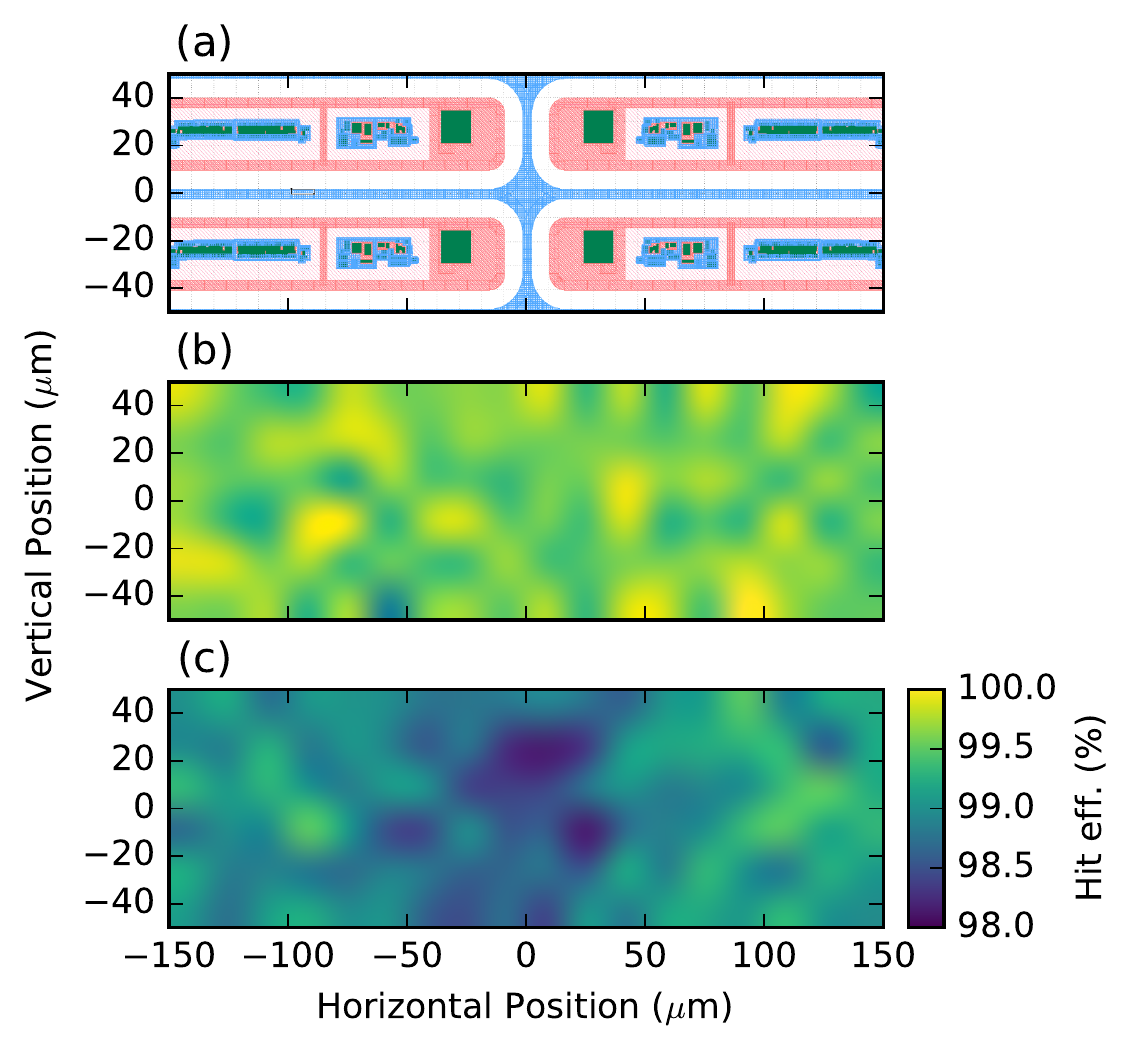}
    \caption{\label{fig:EffInPix} In-pixel efficiency: (a) Layout of four pixels near corners and measured hit detection efficiency maps for (b) un-irradiated and (c) neutron irradiated  LF-Monopix1.
        In (a), n-well, p-well and active area are shown in red, blue, and green respectively.
        The n-wells act as as charge collection nodes and the p-wells 
        separate the collection node from each other.
        In un-irradiated chips (b), the hit detection efficiency is homogeneous throughout the pixels.
        For the irradiated chip in (c) there are only slightly less efficient regions in the regions between pixels.
        Note that due to multiple scattering the resolution is limited. 
    }
\end{figure}

Figure\,\ref{fig:EffInPix} shows the in-pixel efficiency with smaller binning resolution.
The data are superimposed into four pixels in order to increase the statistics.
The measured hit detection efficiency is homogeneous for the un-irradiated chip (see figure\,\ref{fig:EffInPix}(b)).
A small efficiency drop of 1.8\,\% is observed at the corner of pixels for the irradiated chip (see figure\,\ref{fig:EffInPix}(c)).
This can be explained by a decrease of the signal amplitude due to charge trapping, since at the pixel corners the charges are shared between four pixels and the signal charge available for each pixel is correspondingly smaller.

\section{Conclusion}
\label{sec:con}
DMAPS prototypes employing a large electrode design have been developed and characterized in three successive efforts, to establish their suitability for application in high radiation and high hit-rate environments. 
A breakdown voltage of 300~V is realized by the optimization of the guard-ring structure. Bias voltages in excess of 200~V can be applied even after NIEL and TID irradiation to levels of $10^{15}\mathrm{n_{eq}/cm^2}$ and 50\,Mrad, respectively.
TID hardness of the analog front-end circuit has been tested showing no severe performance degradation.
The LF-Monopix1 development achieves a high hit detection efficiency after irradiation of about 99\,\% at a noise occupancy of more than two orders of magnitude lower than the requirement of experiments at the HL-LHC. 

\acknowledgments
We would like to thank W. Snoeys and H. Pernegger for useful discussions
during the process of this development. 
This work has received funding from the European Union’s Seventh Framework and Horizon 2020 research programmes (AIDA-2020 grant 654168 and Marie Curie ITNs TALENT, grant PITNGA-2011-289161, and STREAM, grant 675587) as well as from the German Ministry BMBF, grant 05H15PCA9.


\bibliographystyle{JHEP}
\bibliography{lfpaper_ref}

%
%
%
%
%
%
\end{document}